\documentclass[bm,aps,
amsfonts,amssymb,preprint,nofootinbib]{revtex4}

\usepackage{graphicx}
\usepackage{natbib}

\usepackage{color}

\newcommand\beqa{\begin{eqnarray}}
\newcommand\eeqa{\end{eqnarray}}
\newcommand\n{\nonumber\\}
\begin{document}
{~}

\title{Large Lepton-flavor Mixings from $E_8$ Kodaira Singularity: 
Lopsided Texture via F-theory Family Unification}
\vspace{2cm}
\author{
Shun'ya Mizoguchi\footnote{E-mail: mizoguch@post.kek.jp}
}
\vspace{2cm}
\affiliation{
Theory Center, Institute of Particle and Nuclear Studies, KEK, Tsukuba, Ibaraki, 305-0801, Japan
}


\begin{abstract}
We show that a special type of colliding 7-brane configuration 
of a codimension-two singularity 
realizes not only a 
six-dimensional spectrum with exactly the same quantum 
numbers as that of the three-generation $E_7/(SU(5)\times U(1)^3)$ coset  
family unification model,  but also the three sets of nonchiral singlet
pairs with precisely the correct $U(1)$ charges needed for 
explaining the Yukawa hierarchies and large lepton-flavor mixings 
in a well-known seesaw scenario. 
\end{abstract}


\preprint{KEK-TH 1745}
\date{July 5, 2014}


\maketitle


\section{Introduction}
\label{Introduction}
  {\em Family unification} is a very sophisticated and economical 
 way of understanding the mysterious flavor structures of quarks 
 and leptons. 
 This idea 
hypothesizes that all the quarks 
 and leptons observed in nature are supersymmetric partners 
 of scalars of some supersymmetric coset nonlinear sigma model 
 whose unbroken subgroup includes $SU(5)$. In particular, the coset 
 space $E_7/(SU(5)\times U(1)^3)$ precisely yields \cite{KugoYanagida} 
 three sets of ${\bf \bar 5}\oplus{\bf 10}\oplus {\bf 1}$ of $SU(5)$ besides 
 a single ${\bf 5}$, the former of which is of course to be identified 
 as the quarks and leptons including the right-handed 
neutrinos.  
Thus if we find some mechanism 
that materializes this idea in a more fundamental framework 
such as string theory,  we may get insight into the origin of the 
mysterious structure of the flavors.

In the recent paper \cite{FtheoryFamilyUnification} we have pointed 
out that the charged matter spectrum arising at a codimension-two 
split-type singularity of colliding 7-branes in F-theory  
can be associated with some homogeneous K\"{a}hler manifold 
having the same spectrum, whose defining groups are determined by the 
change of the types of the singularity near the intersection. 
We have also shown there why such a relationship exists by using 
the argument explaining the matter generation in terms of 
string junctions \cite{Tani,CGEH_Three_looks}.  This point of view offers an 
intuitive understanding of matter generation in six dimensions,
and 
allows us to propose 
a special type of colliding 7-brane configuration 
that will realize, after a compactification and a chiral projection, 
the same field content as that of the $E_7/(SU(5)\times U(1)^3)$ 
supersymmetric nonlinear sigma model 
that we mentioned above. 
This is the first step toward realizing the idea of family unification 
in string theory.


In this Letter we further study the brane realization of family unification 
with a special focus on the Yukawa structures. 
%
%
We will show that,
if the enhanced singularity of the coinciding 7-branes is taken to be 
of the $E_8$ type \footnote{by which we mean the type $II^*$ singularity 
in the original classification by Kodaira; the reason for this terminology 
should be obvious.},  then 
the same mechanism may yield, in addition to the three generations 
of matter fields above,  precisely 
the necessary three pairs of Froggatt-Nielsen scalar fields required 
for the explanation of the large 
lepton-flavor mixings proposed by Sato and Yanagida some time ago.
These $SU(5)$ singlet scalar fields are typically charged under the 
anomalous $U(1)$ gauge groups of the model, and naturally expected 
to develop vacuum expectation values due to the FI terms \cite{DSW}. 

The relevance of the $E_8$ singularity to the phenomenological aspects 
of F-theory was pointed out in \cite{E8point,E8point2}. The Froggatt-Nielsen 
mechanism in the $E_8$ GUT was discussed in \cite{FNFtheory}; 
the pattern of charged matter generation is different from ours, however, 
and no reference was made to the coset family unification or Sato-Yanagida's 
idea for deriving the lopsided texture of Yukawa matrices. 

\section{The three-generation $E_7/(SU(5)\times U(1)^3)$ family unification model}
We first very briefly review the family unification based on the four-dimensional 
${\cal N}=1$, $E_7/(SU(5)\times U(1)^3)$ supersymmetric 
nonlinear sigma model.  
In the next section, 
we then turn to the 
argument put forward by Sato and Yanagida \cite{SatoYanagida} 
in an attempt to understand the large lepton-flavor mixings 
in the framework of this coset 
family unification using the Froggatt-Nielsen mechanism. 
For more information on coset family unification,
see \cite{FtheoryFamilyUnification} and references therein.

In order to describe the coset family unification, it is convenient to 
summarize the group theoretical data on $E_7$, and $E_8$, 
which is the key to the ``unification" of the generations of 
charged matter 
{\em and} the Froggatt-Nielsen fields as arising 
from a single geometrical scheme 
in F-theory.

The exceptional Lie algebra $E_8$ 
 is  generated 
by 
traceless 
$E^I_{~~J}$ $ (I,J=1,\ldots,9; ~~I\neq J)$ and antisymmetric tensors  
$E^{IJK}$  and 
$E^*_{IJK}$ $(1\leq I\neq J \neq K \neq I \leq 9)$ 
\footnote{as a complex Lie algebra, 
or a real Lie algebra $E_{8(+8)}$. 
The compact real form of $E_8$ is generated, with real 
coefficients, from a particular set of 
complex linear combinations of these bases (see \cite{MizoguchiYata} 
for explicit expressions).}
if the following commutation relations among them are assumed 
\cite{Freudenthal,MizoguchiE10,MizoguchiGermar}:
%
%
\begin{eqnarray}
\begin{array}{lcl}
{[}E^I_{~J},~E^K_{~L}{]}&=&\delta^{K}_{J} E^I_{~L} -\delta^{I}_{L} E^K_{~J},\\
{[}E^I_{~J},~E^{KLM}{]}&=&3\delta^{[M}_{I}E^{KL]I},\\
{[}E^I_{~J},~E^*_{KLM}{]}&=&-3\delta^{I}_{[M}E^*_{KL]J},\\
{[}E^{IJK},~E^{LMN}{]}&=&-\frac1{3!}\sum_{P,Q,R=1}^9 \epsilon^{IJKLMNPQR}E^*_{PQR},
\\
{[}E^*_{IJK},~E^*_{LMN}{]}&=&+\frac1{3!}\sum_{P,Q,R=1}^9 \epsilon_{IJKLMNPQR}E^{PQR},\\
{[}E^{IJK},~E^*_{LMN}{]}&=&6\delta^J_{[M}\delta_N^K E^I_{~L]}~~~(\mbox{if $I\neq L,M,N$}),\\
{[}E^{IJK},~E^*_{IJK}{]}&=&E^I_{~I}+E^J_{~J}+E^K_{~K}-\frac13\sum_{L=1}^9 E^L_{~L}
\equiv h_{\hat IJK},
\end{array}
\end{eqnarray}
where $\epsilon^{123456789}=\epsilon_{123456789}=+1$.
The numbers of independent components of 
$E^I_{~~J}$, 
$E^{IJK}$  and 
$E^*_{IJK}$ are 80, 84 and 84, each of which consists of 
an irreducible representation of $SL(9)$ of the corresponding dimensions.
Among these generators, the subset: 
\beqa
E^{\hat i}_{~\hat j}~(\hat i \neq \hat j), 
E^{\hat i 89}, E^*_{\hat i 89},
E^{\hat i \hat j \hat k}, E^*_{\hat i \hat j \hat k}, 
h_{\hat i 89}
\label{E7generators}
\eeqa
($\hat i, \hat j, \hat k=1,\ldots,7$), 133 in all,  generate $E_7$.
In the following, we derive various decompositions and $U(1)$ charges 
using this realization of $E_7$. 

In (\ref{E7generators}), we take $E^i_{~j}$ $(i,j=1,\ldots,5)$ (traceless) as 
generators of $SU(5)$
\footnote{For simple notation we abuse the terminology by referring to 
a complex Lie algebra as its compact real form.}. Then their commutant 
(centralizer algebra) 
in $E_7$ is $SU(3)\times U(1)$ generated by 
\beqa
E^6_{~7},E^7_{~6},E^{689},E^{789},E^*_{689},E^*_{789} ,h_{689},h_{789}
\label{SU(3)generators}
\eeqa
and
$10 h_{U(1)_1}+5 h_{U(1)_2}+ 3 h_{U(1)_3}$,
where
\beqa
h_{U(1)_1}&=&2 E^7_{~7}-E^8_{~8}-E^9_{~9},\n
h_{U(1)_2}&=&3 E^6_{~6}-E^7_{~7}-E^8_{~8}-E^9_{~9},\\
h_{U(1)_3}&=&-\frac43\sum_{L=1}^5 E^L_{~L}
+\frac53\sum_{L=6}^9 E^L_{~L}\nonumber
\eeqa
are an orthogonal set of $U(1)$ generators (w.r.t. the Killing form) 
such that the commutant of the $SU(3)$ generated by  (\ref{SU(3)generators}) 
and the $U(1)$ by $h_{U(1)_1}$ is $E_6$, 
the commutant of $h_{U(1)_2}$ in this $E_6$ is $SO(10)$, 
and the commutant of $h_{U(1)_3}$ in this $SO(10)$ is $SU(5)$. 

According to the general rules for extracting the spectrum of the 
supersymmetric coset nonlinear sigma model \cite{BKMU}, 
the $E_7/(SU(5)\times U(1)^3)$ model consists of irreducible 
$SU(5)$ multiplets in the decomposition of $E_7$   
that have negative charges under some 
fixed $U(1)$ group generated by a particular linear combination of 
the generators $h_{U(1)_1}$, $h_{U(1)_2}$ and $h_{U(1)_3}$, 
called ``$Y$-charge" \cite{IKK}. The complex structure of the 
sigma model corresponds one-to-one to the Weyl chamber 
to which the weight vector specified by the $Y$-charge 
belongs \footnote{In F-theory, 
it amounts to the choice of signs of G-fluxes \cite{DonagiWijnholt}.}.
In the present case, if  the generator 
$h_{Y
}$ of the 
$Y$-charge 
is taken 
to be \cite{FtheoryFamilyUnification}
\beqa
h_{Y
}
&=& s h_{U(1)_1} 
+ t(2 h_{U(1)_1} + h_{U(1)_2}) \n
&&+ u (10 h_{U(1)_1} +5h_{U(1)_2} + 3 h_{U(1)_3}) 
\label{Y-charge}
\eeqa 
for some negative $s,t$ and $u$, then 
the $SU(5)$ multiplets corresponding to the $E_7$ generators 
shown in TABLE I have negative $Y$-charges, constituting the spectrum.
\begin{table}
\centering
%
\caption{$U(1)$ charges of the $SU(5)$ multiplets in $E_7/(SU(5)\times U(1)^3)$
 \cite{SatoYanagida}. \label{U(1)charges}}
\begin{tabular}{|c|c|c|c|c|}
\hline 
$SU(5)$ rep.&
$E_7$ generator&
$h_{U(1)_1}$ &
$h_{U(1)_2}$ &
$h_{U(1)_3}$ 
\\
\hline
${\bf 10}_1$
&$E^*_{ijk}$
&$0$
&$0$
&$4$
\\
${\bf 10}_2$
&$E^{ij6}$
&$0$
&$3$
&$-1$
\\
${\bf 10}_3$
&$E^{ij7}$
&$2$
&$-1$
&$-1$
\\
\hline
${\bf \bar 5}_1$
&$E^6_{~i}$
&$0$
&$3$
&$3$
\\
${\bf \bar 5}_2$
&$E^7_{~i}$
&$2$
&$-1$
&$3$
\\
${\bf \bar 5}_3$
&$E^*_{i89}$
&$2$
&$2$
&$-2$
\\
\hline
${\bf 1}_1$
&$E^*_{789}$
&$0$
&$3$
&$-5$
\\
${\bf 1}_2$
&$E^*_{689}$
&$2$
&$-1$
&$-5$
\\
${\bf 1}_3$
&$E^7_{~6}$
&$2$
&$-4$
&$0$
\\
\hline
${\bf 5}$
&$E^{i67}$
&$2$
&$2$
&$2$
\\
\hline 
\end{tabular}
\end{table}
As exhibited in the TABLE I, the spectrum of the supersymmetric 
$E_7/(SU(5)\times U(1)^3)$ nonlinear sigma model consists of 
three sets of ${\bf 10}\oplus {\bar {\bf 5}}\oplus {\bf 1}$ of $SU(5)$ 
and one ${\bf 5}$.
The fermionic components contained in the former are identified 
as the three families of quarks and leptons, and the sigma model 
is assumed to couple to $SU(5)\times U(1)^3$ gauge fields.
The issues of the various anomalies arising from the single {\bf 5} 
will be commented in the final section. The $U(1)^3$ gauge symmetry 
is also anomalous, and will play important roles in the subsequent 
discussions. 

If we set $s=t=0$ in the definition of the $Y$-charge (\ref{Y-charge}), 
then we find that the three singlets become neutral and drop out 
from the spectrum, obtaining the 
$(\bf{10},\bf{\bar 3})\oplus (\bf{\bar 5},\bf{3})\oplus (\bf{5},\bf{1})$
of $SU(5)\times SU(3)$ for the 
original $E_7/(SU(5)\times SU(3) \times U(1))$ Kugo-Yanagida model 
\cite{KugoYanagida}.

\section{Large lepton-flavor mixings}
In \cite{SatoYanagida}, it was postulated 
 that there are three
additional $SU(5)$-singlet complex conjugate pairs of scalar fields 
with a particular assignment of $U(1)$ charges 
in the $E_7/(SU(5)\times U(1)^3)$ model. Let $s_i, \bar s_i$ $(i=0,1,2)$ 
be such scalars, whose $(h_{U(1)_1},h_{U(1)_2},h_{U(1)_3})$ charges are
assumed to be
  \beqa
&&s_0(-3,0,0),s_1(-1,-4,0),s_2(-1,-1,-5),\nonumber\\
&&\bar s_0(+3,0,0),\bar s_1(+1,+4,0),\bar s_2(+1,+1,+5).\label{FGfields}
\eeqa
These artificial-looking assignments are in fact the ones automatically 
realized for the six singlets contained in a {\bf 56} multiplet of $E_7$ \cite{SatoYanagida}, a fact used 
in the geometric realization in the next section.
Following \cite{SatoYanagida}, let us further assume that 
they develop vevs $\langle
s_i
\rangle$
such that
\beqa
\epsilon_i\equiv\frac{\langle
s_i
\rangle}{M_G},~~~
\epsilon_2<<\epsilon_1<< \epsilon_0,
\label{epsilon_assumptions}
\eeqa
where $M_G$ is a high-energy scale not much different 
from the GUT (or the Planck) scale, 
and see the consequences of it.
Denoting a chiral superfield by its $SU(5)$ representation listed 
in TABLE I,  the Yukawa couplings are the coefficients of the 
superpotentials:
\beqa
W_U&=&\sum_{ij}a_{ij} Y_{Uij}
{\bf 10}_i
{\bf 10}_j
{\bf 5}_H,
~W_\nu=\sum_{ij}c_{ij} Y_{\nu ij}
{\bf \bar 5}_i
{\bf 1}_j
{\bf 5}_H,\n
W_D&=&\sum_{ij}b_{ij} Y_{D/Eij}
{\bf \bar 5}_i
{\bf 10}_j
{\bf \bar 5}_H~=~W_E,
\label{Yukawasuperpotentials}
\eeqa
where the Higgs multiplets are 
\beqa
{\bf 5}_H&=&{\bf 5}(2,2,2),\\
{\bf \bar 5}_H&=&\cos\theta~
{\bf \bar 5}_2(2,-1,3)+\sin\theta~
{\bf \bar 5}_3(2,2,-2)
\eeqa
with some angle $\theta$ \footnote{In principle, 
${\bf \bar 5}_1$ could also contribute to ${\bf \bar 5}_H$,  
but in the present case we have made the assumption 
(\ref{epsilon_assumptions}) on the magnitudes of the scalar vevs
so that its contribution would be suppressed and hence is neglected 
here.}. Then up to $O(1)$ factors the Yukawa matrices 
are determined by the requirement for the $U(1)$ charge 
conservations of the superpotentials \cite{FNmechanism,SatoYanagida}:
\beqa
Y_U&\sim&
\mbox{\small$
\left(\begin{array}{ccc}
\epsilon_2^2
&\epsilon_1\epsilon_2
&\epsilon_0\epsilon_2\\
\epsilon_1\epsilon_2
&\epsilon_1^2
&\epsilon_0\epsilon_1\\
\epsilon_0\epsilon_2
&\epsilon_0\epsilon_1
&\epsilon_0^2
\end{array}\right)
$}
,
Y_\nu~\sim~
\mbox{\small$
\left(\begin{array}{ccc}
\epsilon_1^2
&\epsilon_0\epsilon_1
&\epsilon_0\epsilon_2\\
\epsilon_0\epsilon_1
&\epsilon_0^2
&0\\
0&0
&\epsilon_0^2
\end{array}\right)
$},
\nonumber\\
Y_{D/E}&\sim&
\mbox{\small$
\left(\begin{array}{ccc}
\epsilon_1\epsilon_2\cos\theta
&\epsilon_1^2\cos\theta
&\epsilon_0\epsilon_1\cos\theta\\
\epsilon_0\epsilon_2\cos\theta
&\epsilon_0\epsilon_1\cos\theta
&\epsilon_0^2\cos\theta\\
\epsilon_0\epsilon_2\sin\theta
&\epsilon_0\epsilon_1\sin\theta
&\epsilon_0^2\sin\theta
\end{array}\right)
$}
\eeqa
By diagonalizations using (\ref{epsilon_assumptions}), 
we immediately find that  the quark and charged-lepton 
mass ratios are 
\beqa
m_u:m_c:m_t&\sim&
\epsilon_2^2:\epsilon_1^2:\epsilon_0^2\\
m_d:m_s:m_b&\sim&
m_e:m_\mu:m_\tau\\
&\sim
&\epsilon_1\epsilon_2\cos\theta:
\epsilon_0\epsilon_1\sin\theta\cos\theta:
\epsilon_0^2.
\eeqa
Similarly, the Majorana mass matrix for the right-handed neutrinos 
turns out to be
\beqa
M_{\nu_R}&\sim&
M_G
\mbox{\small$
\left(\begin{array}{ccc}
\epsilon_1^2\bar\epsilon_2^2
&\epsilon_0\epsilon_1\bar\epsilon_2^2
&\epsilon_0\epsilon_1\bar\epsilon_1\bar\epsilon_2\\
\epsilon_0\epsilon_1\bar\epsilon_2^2
&\epsilon_0^2\bar\epsilon_2^2
&\epsilon_0^2\bar\epsilon_1\bar\epsilon_2\\
\epsilon_0\epsilon_1\bar\epsilon_1\bar\epsilon_2
&\epsilon_0^2\bar\epsilon_1\bar\epsilon_2
&\epsilon_0^2\bar\epsilon_1^2
\end{array}\right)
$}
.
\eeqa
Therefore, using $m_{\nu_D}\propto Y_\nu$,
the neutrino masses are
\beqa
m_{\nu}&\sim& m_{\nu_D} M_{\nu_R}^{-1}m_{\nu_D}^T
\propto
\mbox{\small$
\left(
\begin{array}{c}\bar\epsilon_1\\
\bar\epsilon_1\\
\bar\epsilon_2
\end{array}
\right)
\left(
\begin{array}{ccc}\bar\epsilon_1&
\bar\epsilon_1&
\bar\epsilon_2
\end{array}
\right)
$}
,
\eeqa
which means a large mixing angle $\theta_{12}$. 
A similar 
analysis 
for $Y_{D/E}Y_{D/E}^T$ shows that the mixing angle 
$\theta_{23}$ is also large.
On the other hand, the CKM matrix is obtained by diagonalizing  
$Y_{D/E}^TY_{D/E}$ and $Y_U^2$, both of which are 
hierarchical. Therefore, the quark mixing angles are small 
in this scenario.

\section{F-theory family unification}
We will now show that the $E_7$ coset structure of the three families 
and the additional three singlet pairs of Froggatt-Nielsen scalars 
in the previous section are 
in fact naturally realized in local F-theory.

The fundamental observation made in \cite{FtheoryFamilyUnification} 
is that the charged matter spectrum of a codimension-two coalesced 
local 7-brane system in F-theory is associated 
one-to-one \footnote{provided that the singularity is of the split type.} 
with a homogeneous K\"{a}hler
 manifold corresponding to the 
change of the type of the singularity near the intersection point.
In the present $E_7/(SU(5)\times U(1)^3)$ case, the matter curve 
\cite{MorrisonVafa,BIKMSV}
is locally 
\cite{FtheoryFamilyUnification}
\beqa
y^2&=&x^3+f(z,w) x + g(z,w),
\label{Weierstrass}
\\
f(z,w)&=&
-3 z^4+z^3+(a \epsilon-3 b^2) z^2+6 b \epsilon^2 z-3 \epsilon^4,
\label{f(z,w)}
\\
g(z,w)&=&2 z^6+\left(\frac{a^2}{12}+3 \epsilon^2+b\right) z^4
+(-2 b^3+a \epsilon b-\epsilon^2) z^3\n
&&+(6 b^2 \epsilon^2-a \epsilon^3) z^2-6 b \epsilon^4 z+2 \epsilon^6,
\label{g(z,w)}
\eeqa
where $a=a(w)$, $b=b(w)$ and $\epsilon=\epsilon(w)$ are 
holomorphic functions 
\footnote{The holomorphy is necessary for supersymmetry \cite{FtheoryFamilyUnification}.} 
only of $w$ such that
$
a(0)=b(0)=\epsilon(0)=0
$.
It describes the 7-brane configuration illustrated in FIG. I (a), where 
the nine 7-branes come to join at $z=w=0$ to develop an $E_7$ 
singularity, but away from that point they become separated 
but only five of them remain on top 
of each other to give an $SU(5)$ singularity. Then the six-dimensional 
hypermultiplets transforming as ${\bf 27}$ of $E_6$ arise from the 
string junctions with an end on the ${\bf A}$ brane bending away 
from the intersecting point, ${\bf 16}$ of $SO(10)$ from the junctions ending 
on one of two 
${\bf C}$ branes, and ${\bf 10}$ of $SU(5)$ from those on the 
${\bf B}$ and the other ${\bf C}$ branes. 
Decomposed into representations of $SU(5)$, they are precisely 
the same set of $SU(5)$ multiplets as those of the $E_7/(SU(5)\times U(1)^3)$ 
sigma model. Therefore, if this six-dimensional theory is further compactified 
on $T^2$ and either chirality of the four-dimensional non-chiral pairs of 
supermultiplets are projected out by an orbifold \cite{Kawamura} 
or turning on appropriate G-fluxes \cite{DonagiWijnholt}, 
one obtains the desired four-dimensional chiral families with 
the $E_7/(SU(5)\times U(1)^3)$ flavor structure.

We now turn to the Froggatt-Nielsen fields. We need three sets of 
$SU(5)$-singlet pairs with $U(1)$ charges (\ref{FGfields});
as we remarked there, they are regarded as coming from 
a ${\bf 56}$ representation of $E_7$. And we note that 
${\bf 56}$ is also 
the representation that constitutes the homogeneous K\"{a}hler
 manifold 
$E_8/(E_7\times U(1))$! Indeed, $E_8$ is decomposed into representations 
of $E_7 \times SU(2)$ as
$
{\bf 248}=({\bf 133},{\bf 1})\oplus
({\bf 56},{\bf 2})\oplus
({\bf 1},{\bf 3})
$,
so the $Y$-charge generator for this coset 
is 
a $U(1)$ in this $SU(2)$, 
and the coset consists of either of the doublet of two ${\bf 56}$s 
and one $E_7$ singlet from an off-diagonal component of $SU(2)$.
Specifically, if we take the $Y$-charge
to be $h_8\equiv E^8_{~8}-E^9_{~9}$,
then we obtain the $SU(5)$ representations 
carrying negative $Y$-charges 
as shown in TABLE II.
\begin{table}
\centering
\caption{The decomposition of $E_8/(E_7\times U(1))$ in 
$SU(5)$ representations.  
All the $SU(5)$ multiplets listed in this table have $h_8$-charge $-1$, 
except the last one $E^9_{~8}$ which has $-2$. They constitute 
a ${\bf 56}$ of $E_7$, except $E^9_{~8}$ which is a singlet.
\label{E8/(E7xU(1))}}
\begin{tabular}{|c|c|c|c|c|}
\hline 
$SU(5)$ rep.&
$E_8$ generator&
$h_{U(1)_1}$ &
$h_{U(1)_2}$ &
$h_{U(1)_3}$ 
\\
\hline

${\bf 1}$($s_0$)
&$E^9_{~7}$
&$-3$
&$0$
&$0$
\\
${\bf 1}$($\bar{s}_0$)
&$E^7_{~8}$
&$+3$
&$0$
&$0$
\\
\hline

${\bf 1}$($s_1$)
&$E^9_{~6}$
&$-1$
&$-4$
&$0$
\\
${\bf 1}$($\bar{s}_1$)
&$E^6_{~8}$
&$+1$
&$+4$
&$0$
\\
\hline 

${\bf 1}$($s_2$)
&$E^*_{678}$
&$-1$
&$-1$
&$-5$
\\
${\bf 1}$($\bar{s}_2$)
&$E^{679}$
&$+1$
&$+1$
&$+5$
\\
\hline 

${\bf 5}$
&$E^i_{~8}$
&$+1$
&$+1$
&$-3$
\\
${\bf \overline{5}}$
&$E^9_{~i}$
&$-1$
&$-1$
&$+3$
\\
\hline 
${\bf 5}$
&$E^{i69}$
&$-1$
&$+2$
&$+2$
\\
${\bf \overline{5}}$
&$E^*_{i68}$
&$+1$
&$-2$
&$-2$
\\
\hline 
${\bf 5}$
&$E^{i79}$
&$+1$
&$-2$
&$+2$
\\
${\bf \overline{5}}$
&$E^*_{i78}$
&$-1$
&$+2$
&$-2$
\\
\hline 
${\bf 10}$
&$E^{ij9}$
&$-1$
&$-1$
&$-1$
\\
${\bf \overline{10}}$
&$E^*_{ij8}$
&$+1$
&$+1$
&$+1$
\\
\hline 

${\bf 1}$
&$E^9_{~8}$
&$0$
&$0$
&$0$
\\
\hline 
\end{tabular}
\end{table}
\\
\indent Thus, in order to include the singlet scalars (\ref{FGfields}) 
in the $E_7/(SU(5)\times U(1)^3)$ model, all we need to do is 
consider the coset $E_8/(SU(5)\times U(1)^4)$  instead,  
where the $Y$-charge for this coset can be chosen to be 
the sum of $h_Y$ (\ref{Y-charge}) and $E^8_{~8}-E^9_{~9}$. 
The $E_8$ coset sigma model has also been studied by many  
authors \cite{Ong,IrieYasui,Buchmuller:1985rc,YanagidaYasui,IKK}.
Of course, if we consider this coset only as a nonlinear sigma model, 
then the scalar couplings would need to contain derivatives and 
the superpotentials (\ref{Yukawasuperpotentials}) would not 
be natural. What is crucial here is that the same spectrum 
can be realized in F-theory, and normally the massless scalars 
in F-theory 
are not considered as Nambu-Goldstone bosons.

\begin{figure}[h]%
\centerline{
\includegraphics[width=0.6\textwidth]{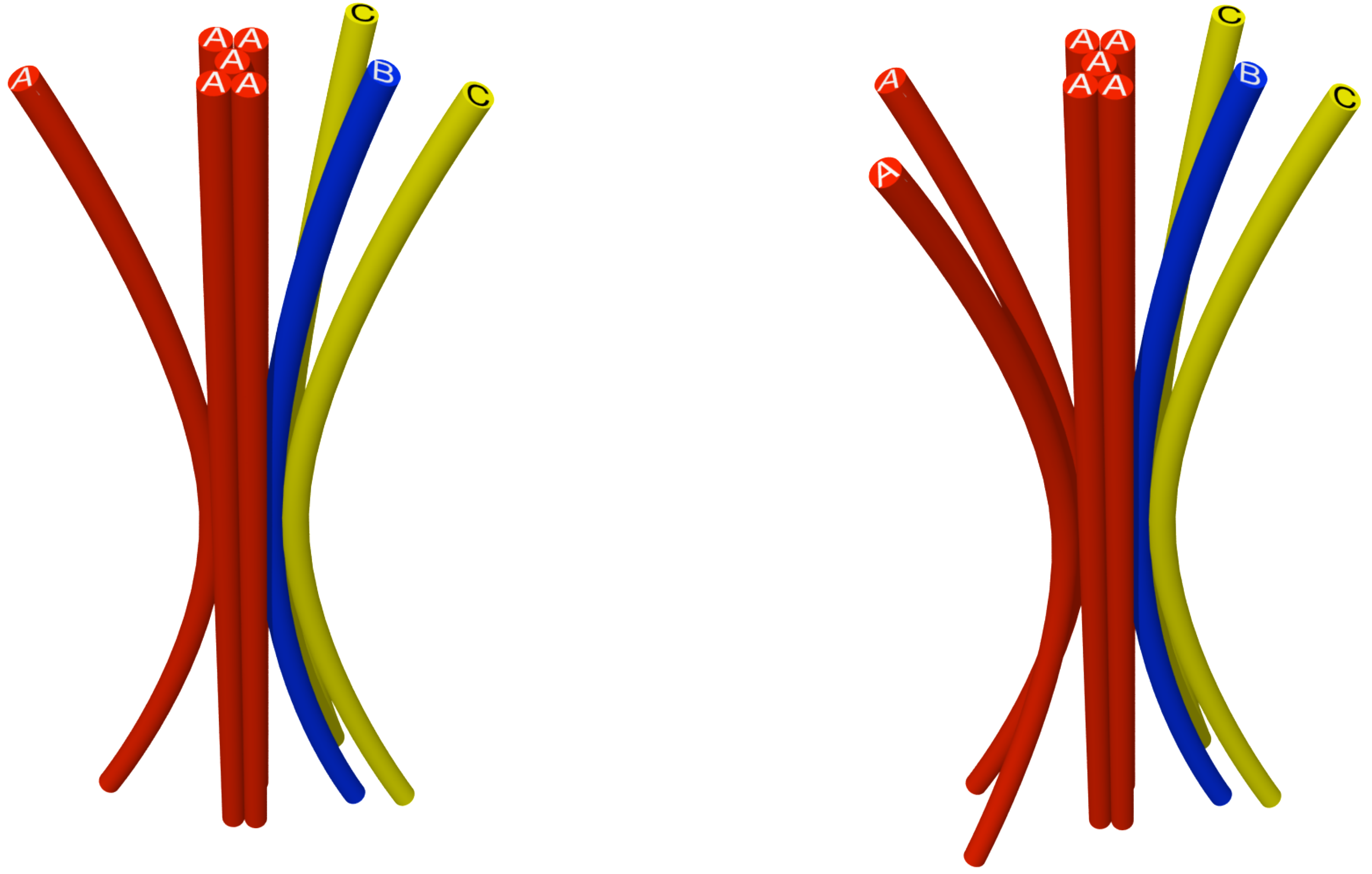}
}
(a)\hspace{35ex}(b)
\caption{\label{}Schematic images of coinciding 7-branes 
corresponding to homogeneous K\"{a}hler
 manifolds:
(a) $E_7/(SU(5)\times U(1)^3)$ (b) $E_8/(SU(5)\times U(1)^4)$. }
\end{figure}

The matter curve corresponding to the 
homogeneous K\"{a}hler
 manifold $E_8/(SU(5)\times U(1)^4)$ is 
represented by the Weierstrass equation (\ref{Weierstrass}) with
\beqa
f(z,w)&=&
-3 z^4+\alpha  z^3+\left(a \epsilon -3 b^2\right) z^2+6 b \epsilon ^2 z-3
   \epsilon ^4,
\label{f(z,w)E8}
\\
g(z,w)&=&
2 z^6+z^5+\left(b \alpha +\frac{a^2}{12}+3 \epsilon ^2\right)
   z^4+(-\epsilon ^2 \alpha -2 b^3 \n && +a \epsilon  b ) z^3
+\left(6 b^2
   \epsilon ^2-a \epsilon ^3\right) z^2-6 b \epsilon ^4 z+2 \epsilon ^6.
\label{g(z,w)E8}
\eeqa
The only differences from the curve with (\ref{f(z,w)}) and (\ref{g(z,w)})
are that it depends on an additional holomorphic parameter 
$\alpha=\alpha(w)$ satisfying 
$
\alpha(0)=0
$, and that $g(z,w)$ (\ref{g(z,w)E8}) contains a $z^5$ term. 
Due to the presence of these terms, the discriminant at $w=0$ reads 
$\Delta=27 z^{10} + 108 z^{11}$,
showing that there are ten 7-branes meeting there to 
exhibit an $E_8$ singularity. The brane configuration is illustrated 
in FIG. I (b). In this case, 
compared to the $E_7$ case,
the string junctions that have an end on the extra {\bf A} brane 
yield the
massless states listed in TABLE II. 
The explicit forms of the string 
junctions can be found in \cite{FtheoryFamilyUnification}. 
\\

\section{Conclusions and discussion}
In this Letter 
we have considered a geometric realization of the idea  
of Sato and Yanagida for explaining the large lepton-flavor 
mixings and hierarchical Yukawa structures 
in local F-theory.
  We have generalized the F-theoretic realization of 
$E_7/(SU(5)\times U(1)^3)$ obtained in \cite{FtheoryFamilyUnification} 
to $E_8/(SU(5)\times U(1)^4)$, which naturally gives rise to
not only three sets of charged matter fields with family non-universality 
but the necessary Froggatt-Nielsen fields from the string junctions 
ending on the extra coinciding 7-brane. 
Although this mechanism alone does not ensure the existence of 
three chiral generations in {\em four} dimensions,  
a further compactification and chiral projection, which may
be implemented by taking an orbifold or turning on G-fluxes, will lead to 
a four-dimensional ${\cal N}=1$ $SU(5)$ GUT. 
If this is done, then we will have an ``all-in-one'' geometric mechanism
in which both the origins of the three families and their large/small 
mixings can be explained in a single setting. Moreover, the singlet scalars 
are charged under anomalous $U(1)$s and an FI term will be generated, 
leading to their acquiring nonzero vevs and triggering  
SUSY breaking \cite{DvaliPomarol,BinetruyDudas} and other 
effects (see e.g. \cite{KNTY}).

Though interesting, however, the following issues must be explored 
in order for this model to be considered as a realistic model: 
(i) How the $SU(5)$ anomaly cancels  
(ii) How the FN fields get the sizes of vevs (\ref{epsilon_assumptions})
(iii) How it can be embedded in a global Calabi-Yau and how 
the neutral moduli are stabilized (iv) How such a brane collision 
comes into being dynamically. Possible solutions for some of these 
issues have been suggested in \cite{FtheoryFamilyUnification}. 
As for (i), a possible origin of an extra ${\bf \bar 5}$ to compensate 
the anomaly \cite{YanagidaYasui} is one emerging from the 
orbifold fixed points since, in the heterotic dual picture (if available), 
the twisted sector would automatically cure the lack of modular invariance,  
which is believed to be equivalent to gauge invariance (see e.g. \cite{BSS}). 
We hope to report on these issues elsewhere.
\\

The author thanks T.~Kobayashi and Y.~Yasui for useful discussions.
This work is supported by Grant-in-Aid
for Scientific Research  
(C) \#25400285 
 and (A) \#26247042 from
The Ministry of Education, Culture, Sports, Science
and Technology of Japan.




\begin{thebibliography}{00}



\bibitem{KugoYanagida} 
  T.~Kugo and T.~Yanagida,
  Phys.\ Lett.\ B {\bf 134}, 313 (1984).


\bibitem[Mizoguchi(2014)]{FtheoryFamilyUnification}S.~Mizoguchi, 
F-theory Family Unification, arXiv: 1403. 7066 [hep-th]. To appear in JHEP.

\bibitem[Tani(2001)]{Tani}T.~Tani
T.~Tani,
  Nucl.\ Phys.\ B {\bf 602}, 434 (2001).

\bibitem{CGEH_Three_looks} 
M.~Cvetic, I.~Garcia Etxebarria and J.~Halverson,
  JHEP {\bf 1111}, 101 (2011)
  [arXiv:1107.2388 [hep-th]].

\bibitem{DSW}
M.~Dine, N.~Seiberg and E.~Witten,
  Nucl.\ Phys.\ B {\bf 289}, 589 (1987).

\bibitem{E8point}
J.~J.~Heckman, A.~Tavanfar and C.~Vafa,
  JHEP {\bf 1008}, 040 (2010)
  [arXiv:0906.0581 [hep-th]].

\bibitem{E8point2}
J.~Marsano, N.~Saulina and S.~Schafer-Nameki,
  JHEP {\bf 0908}, 046 (2009)
  [arXiv:0906.4672 [hep-th]].

\bibitem{FNFtheory}
E.~Dudas and E.~Palti,
  JHEP {\bf 1001}, 127 (2010)
  [arXiv:0912.0853 [hep-th]].


\bibitem{SatoYanagida}
J.~Sato and T.~Yanagida,
  Phys.\ Lett.\ B {\bf 430}, 127 (1998)
  [hep-ph/9710516].
  
  
\bibitem{MizoguchiYata}
  S.~Mizoguchi and M.~Yata,
  PTEP {\bf 2013}, no. 5, 053B01 (2013)
  [arXiv:1211.6135 [hep-th]].
  
  
\bibitem{Freudenthal}
H. Freudenthal, Proc. Kon. Ned. Akad. Wet. A56 (Indagationes Math. 15)
(1953) 95-98 (French).

\bibitem{MizoguchiE10} 
  S.~Mizoguchi,
  Nucl.\ Phys.\ B {\bf 528}, 238 (1998)
  [hep-th/9703160].

\bibitem{MizoguchiGermar} 
S.~Mizoguchi and G.~Schroder,
  Class.\ Quant.\ Grav.\  {\bf 17}, 835 (2000)
  [hep-th/9909150].

\bibitem{DonagiWijnholt}
R.~Donagi and M.~Wijnholt,
  Adv.\ Theor.\ Math.\ Phys.\  {\bf 15}, 1237 (2011)
  [arXiv:0802.2969 [hep-th]].

\bibitem{IKK}
K.~Itoh, T.~Kugo and H.~Kunitomo,
ian Coset Space G/h,''
  Nucl.\ Phys.\ B {\bf 263}, 295 (1986);
ian Coset Spaces G/h: G = E6, E7 And E8,''
  Prog.\ Theor.\ Phys.\  {\bf 75}, 386 (1986).

\bibitem{BKMU} 
  M.~Bando, T.~Kuramoto, T.~Maskawa and S.~Uehara,
  Phys.\ Lett.\ B {\bf 138}, 94 (1984);
  Prog.\ Theor.\ Phys.\  {\bf 72}, 313 (1984);
  Prog.\ Theor.\ Phys.\  {\bf 72}, 1207 (1984).

\bibitem{FNmechanism}
C.~D.~Froggatt and H.~B.~Nielsen,
  Nucl.\ Phys.\ B {\bf 147}, 277 (1979).

\bibitem{MorrisonVafa}
D.~R.~Morrison and C.~Vafa,
  Nucl.\ Phys.\ B {\bf 473}, 74 (1996)
  [hep-th/9602114];
  %
  Nucl.\ Phys.\ B {\bf 476}, 437 (1996)
  [hep-th/9603161].


\bibitem{BIKMSV}
M.~Bershadsky, K.~A.~Intriligator, S.~Kachru, D.~R.~Morrison, V.~Sadov and C.~Vafa,
  Nucl.\ Phys.\ B {\bf 481}, 215 (1996)
  [hep-th/9605200].


\bibitem{Kawamura}
Y.~Kawamura,
Prog.\ Theor.\ Phys.\  {\bf 103}, 613 (2000)
[arXiv:hep-ph/9902423];
Prog.\ Theor.\ Phys.\  {\bf 105}, 691 (2001)
[arXiv:hep-ph/0012352];
Prog.\ Theor.\ Phys.\  {\bf 105}, 999 (2001)
[arXiv:hep-ph/0012125].



\bibitem{Ong} 
C.~-L.~Ong,
  Phys.\ Rev.\ D {\bf 27}, 3044 (1983);
%
  Phys.\ Rev.\ D {\bf 31}, 3271 (1985).

\bibitem{IrieYasui}
S.~Iri\'{e} and Y.~Yasui, Z.\ Phys.\ C {\bf 29} (1985), 123.

\bibitem{Buchmuller:1985rc} 
  W.~Buchmuller and O.~Napoly,
  Phys.\ Lett.\ B {\bf 163}, 161 (1985).

\bibitem{YanagidaYasui} 
  T.~Yanagida and Y.~Yasui,
  Nucl.\ Phys.\ B {\bf 269}, 575 (1986).








\bibitem{DvaliPomarol}  
G.~R.~Dvali and A.~Pomarol,
  Phys.\ Rev.\ Lett.\  {\bf 77}, 3728 (1996)
  [hep-ph/9607383].

\bibitem{BinetruyDudas} 
P.~Binetruy and E.~Dudas,
  Phys.\ Lett.\ B {\bf 389}, 503 (1996)
  [hep-th/9607172].

\bibitem{KNTY}
T.~Kobayashi, H.~Nakano, H.~Terao and K.~Yoshioka,
  Prog.\ Theor.\ Phys.\  {\bf 110}, 247 (2003)
  [hep-ph/0211347].

\bibitem{BSS}
T.~Banks, N.~Seiberg and E.~Silverstein,
  Phys.\ Lett.\ B {\bf 401}, 30 (1997)
  [hep-th/9703052].


\end{thebibliography}


\end{document}